# "A more probable explanation" is still impossible to explain GN-z11-flash: in response to Steinhardt et al. (arXiv:2101.12738)

Linhua Jiang[1,2], Shu Wang[1,2], Bing Zhang[3], Nobunari Kashikawa[4,5], Luis C. Ho[1,2], Zheng Cai[6], Eiichi Egami[7], Gregory Walth[8], Yi-Si Yang[9,10], Bin-Bin Zhang[9,10], Hai-Bin Zhao[11]

*[1]Kavli Institute for Astronomy and Astrophysics, Peking University, Beijing, China*

*[2]Department of Astronomy, School of Physics, Peking University, Beijing, China*

*[3]Department of Physics and Astronomy, University of Nevada, Las Vegas, NV, USA*

*[4]Department of Astronomy, Graduate School of Science, The University of Tokyo, Tokyo, Japan*

*[5]Optical and Infrared Astronomy Division, National Astronomical Observatory, Tokyo, Japan*

*[6]Department of Astronomy, Tsinghua University, Beijing, China*

*[7]Steward Observatory, University of Arizona, Tucson, AZ, USA*

*[8]Observatories of the Carnegie Institution for Science, Pasadena, CA, USA*

*[9]School of Astronomy and Space Science, Nanjing University, Nanjing, China*

*[10]Key Laboratory of Modern Astronomy and Astrophysics (Nanjing University), Ministry of Education, China*

*[11]CAS Key Laboratory of Planetary Sciences, Purple Mountain Observatory, Chinese Academy of Sciences, Nanjing, China*

## ABSTRACT

In Jiang et al. (2020), we reported a possible bright flash (hereafter GN-z11-flash) from a galaxy GN-z11 at $z \approx 11$. Recently, Steinhardt et al. (2021; arXiv:2101.12738) found 27 images with transient signals in Keck MOSFIRE archival data and claimed that GN-z11-flash was more likely from a moving object in our Solar system. We show that the Steinhardt et al.'s definition of the chance probability and their methodology of finding GN-z11-flash-like transients are problematic in several aspects. In particular, none of their transients is analogous to GN-z11-flash, and none of them is positionally coincident with a known object in their imaging data. In Jiang et al., we performed a comprehensive analysis of the origin of GN-z11-flash and ruled out, to the best of our knowledge, the possibility of known man-made objects or moving objects in the Solar system, based on all available information and our current understanding of these objects. Steinhardt et al. did not use such information and did not analyse the GN-z11-flash event itself. The majority of their transients are apparently low-Earth orbit satellites or aircrafts. Therefore, their analysis can neither prove nor disprove our results. Finally, we present a method to estimate the chance probability of finding GN-z11-flash-like transients in archival data. Based on this method and the archival data used by Steinhardt et al., we obtain a loose upper limit of the probability that actually supports the original results of Jiang et al. (2020).

# 1. INTRODUCTION

In 2017 April, we carried out spectroscopic observations of GN-z11 using Keck MOSFIRE and obtained 106 images in the *K* band. In one of the individual exposures, we detected a bright flash, a compact continuum emission with a spatial position coincident with the position of the GN-z11 emission lines that we identified. We then performed a comprehensive analysis of the origin of this flash (hereafter referred to as GN-z11-flash) and ruled out, to the best of our knowledge, the possibility of known man-made objects or moving objects in the Solar system, based on the observational information and our current understanding of the properties of these objects. We found that GN-z11-flash can be explained by a rest-frame UV flash associated with a long gamma-ray burst (GRB) from GN-z11. Detailed information is described in Jiang et al. (2020). Another interpretation was proposed by Padmanabhan & Loeb (2021), who found that GN-z11-flash was likely from a shock-breakout of a Pop III supernova explosion occurring in GN-z11.

Recently, Steinhardt et al. (2021; arXiv:2101.12738) searched 12,300 Keck MOSFIRE images in the archive and found 27 single-exposure transients. They defined their chance probability of finding a transient like GN-z11-flash as the fraction of the images with transients. They then compared this fraction with our chance probabilities for GN-z11-flash and claimed that GN-z11-flash was more likely from an Earth-orbiting satellite, without analysing the GN-z11-flash event itself. As mentioned above, our chance probabilities were carefully estimated and the probability of known objects was ruled out, by considering all available information of GN-z11-flash and the properties of the known objects. Steinhardt et al. did not consider such information, although they have suspected that their transients were likely Earth-orbiting satellites. As shown below, the majority of them were likely low-Earth orbit satellites or aircrafts. Therefore, the transients found by Steinhardt et al. are not GN-z11-flash-like transients.

# 2. ANALYSIS

First, the definition of the chance probability of finding a transient by Steinhardt et al. is inappropriate. Steinhardt et al. found 27 images with transient signals, including eight images in GOODS-N (excluding GN-z11-flash), in 12,300 images that they have checked. Their chance probability was calculated as 27/12,300. Let us assume a slightly extreme case in which MOSFIRE has a field-of-view (FoV) of 5 deg$^2$. From a quick calculation considering the current FoV of $3'\times6'$, one would see one transient in each exposure on average (in the dawn or dusk hours, see our analysis later). Based on the definition by Steinhardt et al., the chance probability would be 100%. This simple fraction is certainly not the chance probability that we calculated for GN-z11-flash. Our chance probability is the probability that one detects an event like GN-z11-flash in our target region of 0.9" × 0.7", or the probability that one object moves across this region, after we rule out the probability of all known sources. Here 0.9" is the slit width and 0.7" is our tolerance of the position uncertainty along the slit (roughly four pixels).

Second, the transients found by Steinhardt et al. are not GN-z11-flash-like transients. Steinhardt et al. included all transients that they identified, although nearly all of them are likely man-made satellites or aircrafts. In Jiang et al., we performed a comprehensive analysis of GN-z11-flash and ruled out known sources using all available information, including observational information such as date, time, site location, target position (its R.A. and Dec.), slit mask design (e.g., slit position angle and target position), spectral properties (spectral slope and brightness), and the properties (orbit, brightness, and angular velocity) of man-made satellites and moving objects in the Solar system. We also checked satellite databases and the IAU Minor Planet Center database. Steinhardt et al. did not use such information. As Jiang et al. explained, such information is critical. For example, GN-z11-flash was detected 3.5 hours after sunset, and GN-z11 was rising in the eastern sky. At that moment, for an observer at the observing site (Mauna Kea at latitude ~20º north) toward the direction of GN-z11, a low-Earth or medium-Earth orbit satellite with height <4000 km would not have been able to reflect the light of the Sun. This immediately rules out low-Earth orbit satellites; because satellites reflect sunlight, low-Earth orbit satellites are nearly invisible or very faint during the night. In fact, the majority of the transients listed by Steinhardt et al. were detected in the dawn or dusk hours (see later).

Third, another critical difference between GN-z11-flash and the transients identified by Steinhardt et al. is that GN-z11-flash is positionally associated with the known galaxy GN-z11, while other transients are not. When a transient does not have a corresponding object in the extremely deep imaging data, the explanation of a moving object in our Solar system becomes natural. This makes GN-z11-flash completely different from the transients found by Steinhardt et al. Therefore, if one really wants to estimate the chance probability for GN-z11-flash-like transients from the MOSFIRE data, one must select transients that were positionally associated with known objects in imaging data.

Finally, Steinhardt et al. did not analyze the GN-z11-flash event itself, despite their stated aim of doing so. As mentioned above, the available information is essential to estimate the chance probability for transient events such as GN-z11-flash. However, Steinhardt et al. chose to use a simple fraction to infer the chance probability. There are always concerns when applying a generic description to a specific individual if one does not perform a careful analysis. For example, the existence of many artificial radio burst signals from the Parkes 64m radio telescopes (the so-called "perytons" that were eventually identified as artificial signals from microwave ovens; Petroff et al. 2015) had for some time led to a wrong impression that there are no genuine cosmological radio bursts. However, the detailed analysis of the photo-type of these bursts (Lorimer et al. 2007) revealed unique characteristics that are not shared by perytons. We now know that they (known as fast radio bursts) are indeed of an astrophysical origin.

We looked into eight transients (excluding GN-z11-flash) identified by Steinhardt et al. in GOODS-N. We chose this field because this is a famous deep field at high Galactic latitude without bright stars and far away from the plane of the Ecliptic. We found that six of them were observed in the dawn hours. Two of the six also show very extended spectral features (MF.20130215.56333 and MF.20160102.57635; MF.20130215.56333 is shown in the upper panel of Figure 1). This suggests that these six transients were highly likely low-Earth orbit satellites or aircrafts. Another one also shows an extended spectral feature (MF.20170304.45898; lower panel of Figure 1), suggesting an aircraft origin. There is only one transient in this sample that appears point-like and was observed during midnight. Since its 2D spectrum is noisy, and it is not associated with a known object in the imaging data, we did not analyse it in detail. In summary, we are not convinced that Steinhardt et al.'s sample contains GN-z11-flash-like transients.

### 3. A METHOD TO ESTIMATE CHANCE PROBABILITY

Jiang et al. (2020) ruled out, to the best of their knowledge, the possibility that GN-z11-flash was from known sources and proposed that the flash was likely a rest-frame UV flash associated with a long GRB from GN-z11. In principle, one can use archival multi-object spectral data to estimate the chance probability for GN-z11-flash-like events and compare it with our results. However, this must be done in a very rigorous way. We briefly demonstrate our own method in three steps: (1) identify transients in the archival data; (2) identify transients that have corresponding counterparts in imaging data; (3) perform a rigorous analysis to rule out or estimate the probability of all types of known objects mentioned in Jiang et al. (2020). One may switch the first two steps, i.e., first is to identify target positions, and then search for transient signals at these positions.

We apply the above method to analyse Steinhardt et al.'s archival data and estimate a loose upper limit for the chance probability that a GN-z11-flash-like object passes across a slit at a given position in one exposure. The tolerance of the position uncertainty is still 0.7". There are 6000 images, and the MOSFIRE FoV is $3' \times 6'$, where $6'$ is along the spatial direction. We ignore gaps along the spatial direction in slit masks. As mentioned, we did not find any GN-z11-flash-like objects in GOODS-N based on the information provided by Steinhardt et al. The upper limit is then $1/[6000 \times 6 \times (60/0.7)] \sim 3 \times 10^{-7}$. This supports the conclusions of Jiang et al. (2020).

### 4. CONCLUSION

We have shown that the definition of the chance probability of finding GN-z11-flash-like transients by Steinhardt et al. is inappropriate and is completely different from that of Jiang et al. (2020). We have also shown that the Steinhardt et al.'s methodology of finding GN-z11-flash-like transients is questionable in several aspects. None of their transients is like GN-z11-flash. We further applied our method to estimate the chance probability of finding GN-z11-flash-like transients to the archival data used by Steinhardt et al. The results of this analysis actually support the conclusions of Jiang et al. (2020).

**Acknowledgements**

We acknowledge support from the National Science Foundation of China (11721303, 11890693, 11991052) and the National Key R&D Program of China (2016YFA0400702, 2016YFA0400703). We thank the authors of Steinhardt et al. (2021) for their paper and helpful discussion.

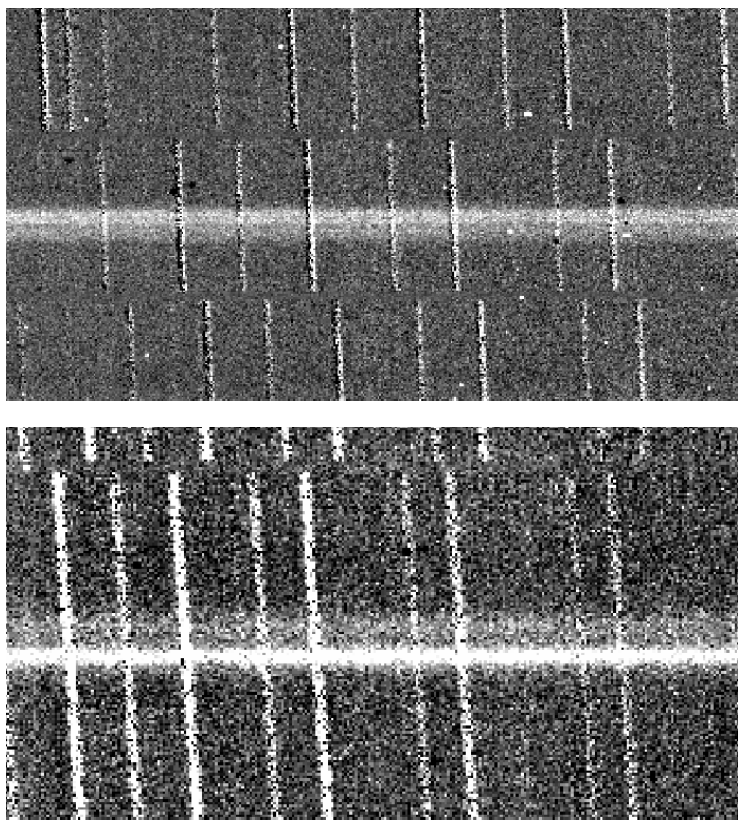

Figure 1. Part of the spectral images of two transients from Steinhardt et al. (2021). Upper panel: MF.20130215.56333. Lower panel: MF.20170304.45898. Both sources are extended, suggesting that they are likely nearby objects.